\documentclass[prd,aps,twocolumn,preprintnumbers,showpacs, nofootinbib,superscriptaddress,notitlepage]{revtex4-1}

\usepackage{graphicx,epsfig,mathtools}
\usepackage[usenames]{color}
\usepackage{slashed}
\usepackage{epstopdf}
\usepackage{multirow}

\allowdisplaybreaks

\begin{document}
	
\title{Baryons in the light-front approach: The three-quark picture}
	
\author{Zhen-Xing Zhao}\email{Email:zhaozx19@imu.edu.cn}
\affiliation{School of Physical Science and Technology, Inner Mongolia University, Hohhot 010021, China}

\author{Fu-Wei Zhang}
\affiliation{School of Physical Science and Technology, Inner Mongolia University, Hohhot 010021, China}

\author{Xiao-Hui Hu}\email{Email:huxiaohui@cumt.edu.cn}
\affiliation{School of Materials Science and Physics, China University of Mining and Technology,  Xuzhou 221116, China}

\author{Yu-Ji Shi}\email{Email:shiyuji@ecust.edu.cn}
\affiliation{School of Physics, East China University of Science and Technology, Shanghai 200237, China}

\begin{abstract}
In this work, a three-quark picture is constructed using a bottom-up approach for baryons in light-front quark model. The shape parameters, which characterize the momentum distribution inside a baryon, are determined with the help of the pole residue of the baryon. The relation between the three-quark picture and the diquark picture is clarified. When building the model, we find that Lorentz boost plays a crucial role, and the bottom-up modeling approach can be generalized to multiquark states. Based on this, a unified theoretical framework for describing multiquark states may be established. As a by-product of model construction, we can easily obtain a newly improved definition of baryon interpolating current. The hadron interpolating currents are the starting point of Lattice QCD and QCD sum rules, and therefore are of great importance.
\end{abstract}

\maketitle

\section{Introduction}

Recently, there have been some big events in the field of heavy flavor physics, including the discovery of CP violation (CPV) in neutral charm mesons~\cite{LHCb:2019hro} and the discovery of doubly charmed baryons~\cite{LHCb:2017iph}. On the one hand, CPV has been confirmed in many meson systems~\cite{Belle:2001zzw,BaBar:2001pki,Christenson:1964fg},
however, CPV has never been observed in the baryon sector and to search
for baryonic CPV is becoming particularly urgent~\cite{Wang:2022fih}.
On the other hand, the discovery of doubly charmed baryon has attracted
great interests both theoretically and experimentally. The study of the properties
of heavy baryons plays an important role in accurately testing the
standard model, searching for the origin of CP violation and new physics,
and understanding strong interactions.

From a theoretical perspective, baryons are generally more complicated than mesons. In spite of this, there have been many
methods to study the decay properties of heavy flavor baryons, including
the light-front quark model, SU(3) flavor symmetry, effective field
theory, QCD and light-cone sum rules, perturbative QCD, Lattice
QCD, and so forth. Some recent progress can be found in Refs.~\cite{Wang:2017mqp,Zhao:2018zcb,Zhao:2018mrg,Xing:2018lre,Zhao:2022vfr,Wang:2022ias,Liu:2022mxv,Wang:2017azm,He:2018joe,Zhang:2018llc,Shi:2020qde,Qin:2021zqx,Shi:2019hbf,Xing:2021enr,Zhao:2020mod,Zhao:2021sje,Shi:2019fph,Shi:2022kfa,Shi:2022zzh,Han:2022srw,Zhang:2021oja}.

In Ref.~\cite{Wang:2017mqp}, light-front quark model (LFQM) is
used to investigate the weak decays of doubly heavy baryons under
the diquark picture. In this picture, the two quarks which do not
participate in weak interactions are considered to form a loosely bound system -- a diquark.
In this way, the relatively complicated three-body problem inside
a baryon is reduced to a relatively simple two-body problem. However,
this diquark picture is criticized by some people. Take the weak decays
of $\Xi_{bc}(bcq)$ as an example. When the $b$ quark decays, $(cq)$
is considered as a diquark, while when the $c$ quark decays, $(bq)$ is
considered as a diquark. From this, one can see that the diquark
picture is actually a matter of expediency. In addition, the diquark
picture inevitably contains more parameters, such as the diquark masses.
Even for the $0^{+}$ diquark $[ud]$ and the $1^{+}$ diquark $\{ud\}$
containing the same quark components, their masses are considered
to be different~\cite{Jaffe:2004ph}.

Under the diquark picture, in Ref.~\cite{Ke:2012wa}, the baryon-quark-diquark
vertex is given by
\begin{equation}
-\frac{1}{\sqrt{3}}\gamma_{5}\slashed\epsilon^{*}(p_{2},\lambda_{2}),
\end{equation}
while in Ref.~\cite{Chua:2018lfa}, it is
\begin{align}
 & \frac{1}{\sqrt{3}}\gamma_{5}\slashed\epsilon^{*}(\bar{P},\lambda_{2})\nonumber \\
= & \frac{1}{\sqrt{3}}\gamma_{5}\left(\slashed\epsilon^{*}(p_{2},\lambda_{2})-\frac{M_{0}+m_{1}+m_{2}}{p_{2}\cdot\bar{P}+m_{2}M_{0}}\epsilon^{*}(p_{2},\lambda_{2})\cdot\bar{P}\right),
\end{align}
where $p_{1,2}$ ($m_{1,2}$) are respectively the momenta (masses) of quark 1 and the
diquark, $\lambda_{2}$ ($\epsilon$)
is the helicity (polarization vector) of the diquark, $\bar{P}\equiv p_{1}+p_{2}$
and $M_{0}^{2}\equiv\bar{P}^{2}$. In fact, except for an unimportant
negative sign, there is a Lorentz boost between the two expressions,
and the latter have correctly considered this effect. Later, one can 
see that this Lorentz boost effect plays a crucial role in our model
construction.

Early in 1998, the authors of Ref.~\cite{Tawfiq:1998nk} developed the
three-quark picture of heavy baryons. Recently, the three-quark picture
has been used to study the weak decays of heavy flavor baryons in
Refs.~\cite{Ke:2019smy,Lu:2023rmq,Geng:2021nkl,Geng:2022xpn,Li:2021kfb,Li:2022hcn}, and this
work aims to highlight the following points: 
\begin{itemize}
\item Spin wavefunction will be constructed in a bottom-up approach. In this
work, we are limited to considering ground state baryons. Several
typical baryons include: $\Lambda_{Q}$, $\Sigma_{Q}$, and $\Sigma_{Q}^{*}$.
This is due to the following coupling in spin space: $\left(\frac{1}{2}\otimes\frac{1}{2}\right)\otimes\frac{1}{2}=(0\oplus1)\otimes\frac{1}{2}=\frac{1}{2}\oplus\frac{1}{2}\oplus\frac{3}{2}.$
One of our potentially important discoveries is
that Lorentz boost may become more important when constructing
the quark model of multiquark states, which have a certain size, and
are multi-body QCD bound states. Since we may still lack a quark model
that can describe all the multiquark states, this discovery may be
a key to unlock the door. It is worth noting that LFQM has been applied to
some relatively simple multiquark states in Refs.~\cite{Cheng:2004cc,Cheng:2004ew}.
In addition, as a by-product of model construction,
we can easily obtain improved definitions of interpolating
currents of baryons.
\item A method for determining the shape parameters will be proposed. The
shape parameters (see Eq.~(\ref{eq:momentum_wf})) characterize the
momentum distribution inside a hadron. For the meson case, the shape
parameter is determined by the decay constant of the meson~\cite{Cheng:2003sm}.
The ``decay constant" of a baryon is the
pole residue, which is used to determine the shape parameters
of the baryon, see below.
\item The relation between the diquark picture and the three-quark picture
will be clarified. In this work, we will consider three weak decays:
$\Lambda_{b}\to\Lambda_{c}$, $\Sigma_{b}\to\Sigma_{c}$, and $\Xi_{cc}\to\Lambda_{c}$.
For the first two processes, the spectator quarks are respectively
a scalar and axial-vector diquark, while for the last process, a diquark
is broken up in the initial state and a new diquark emerges by rearranging
quarks in the final state. Here, for $\Xi_{cc}\to\Lambda_{c}$, the
two charmed quarks in $\Xi_{cc}$ are usually considered as an axial-vector
diquark and the $u,d$ quarks in $\Lambda_{c}$ form a scalar
one. In the diquark picture, overlap factors are important quantities for obtaining 
the physical form factors~\cite{Wang:2017mqp}. We will derive the overlap factors
for $\Xi_{cc}\to\Lambda_{c}$, through which the relation between
the diquark picture and the three-quark picture can be illustrated.
\end{itemize}

The rest of this article is arranged as follows. In Sec.~II, theoretical
framework and some applications are introduced, including: the definitions
of baryon states; the determination of the shape parameters; the form
factors of $\Lambda_{b}\to\Lambda_{c}$, $\Sigma_{b}\to\Sigma_{c}$,
and $\Xi_{cc}\to\Lambda_{c}$; the relation between the two pictures;
improved definitions of interpolating currents of baryons.
In Sec.~III, numerical results of shape parameters, form factors,
and semileptonic decay widths will be shown and will be compared with
others in the literature. We conclude this article in the last section. 

\section{Theoretical framework and some applications}

\subsection{The baryon states}

\label{subsec:The-baryon-states}

In this section, we will consider three baryon states: $\Lambda_{Q}$,
$\Sigma_{Q}$ and $\Sigma_{Q}^{*}$. They all have the same quark components
$udQ$ and are all S-wave baryons, and their spins are respectively
$1/2$, $1/2$, and $3/2$.

Under the three-quark picture, the baryon state in LFQM
is expressed as 
\begin{widetext}
\begin{eqnarray}
|{\cal B}(P,S,S_{z})\rangle & = & \int\{d^{3}\tilde{p}_{1}\}\{d^{3}\tilde{p}_{2}\}\{d^{3}\tilde{p}_{3}\}2(2\pi)^{3}\delta^{3}(\tilde{P}-\tilde{p}_{1}-\tilde{p}_{2}-\tilde{p}_{3})\frac{1}{\sqrt{P^{+}}}\nonumber \\
 & \times & \sum_{\lambda_{1},\lambda_{2},\lambda_{3}}\Psi^{SS_{z}}(\tilde{p}_{1},\tilde{p}_{2},\tilde{p}_{3},\lambda_{1},\lambda_{2},\lambda_{3})C^{ijk}|q_{1}^{i}(p_{1},\lambda_{1})q_{2}^{j}(p_{2},\lambda_{2})q_{3}^{k}(p_{3},\lambda_{3})\rangle,\label{eq:baryon_state}
\end{eqnarray}
\end{widetext}
where $p_{i}$ ($\lambda_{i}$) is the light-front momentum (helicity)
of the $i$-th quark, the color wavefunction $C^{ijk}=\epsilon^{ijk}/\sqrt{6}$, and
the spin and momentum wavefunctions are contained in $\Psi^{SS_{z}}$. The light-front
momentum is decomposed into $p_{i}=(p_{i}^{-},p_{i}^{+},p_{i\perp})$
and
\begin{align}
 & \tilde{p}_{i}=(p_{i}^{+},p_{i\perp}),\quad p_{i\perp}=(p_{i}^{1},p_{i}^{2}),\quad p_{i}^{-}=\frac{m_{i}^{2}+p_{i\perp}^{2}}{p_{i}^{+}},\nonumber \\
 & \{d^{3}\tilde{p}_{i}\}=\frac{dp_{i}^{+}d^{2}p_{i\perp}}{2(2\pi)^{3}}.
\end{align}
The intrinsic variables $(x_{i},k_{i\perp})$ are introduced through
\begin{align}
 & p_{i}^{+}=x_{i}P^{+},\quad p_{i\perp}=x_{i}P_{\perp}+k_{i\perp},\nonumber \\
 & \sum_{i=1}^{3}x_{i}=1,\quad\sum_{i=1}^{3}k_{i\perp}=0,
\end{align}
where $x_{i}$ is the light-front momentum fraction constrained by
$0<x_{i}<1$. The invariant mass $M_{0}$ is defined by $M_{0}^{2}\equiv\bar{P}^{2}$
with $\bar{P}=p_{1}+p_{2}+p_{3}$, and it can be shown that
\begin{equation}
M_{0}^{2}=\frac{k_{1\perp}^{2}+m_{1}^{2}}{x_{1}}+\frac{k_{2\perp}^{2}+m_{2}^{2}}{x_{2}}+\frac{k_{3\perp}^{2}+m_{3}^{2}}{x_{3}}.
\end{equation}
$M_{0}$ is in general different from the baryon mass $M$ which obeys
the condition $M^{2}=P^{2}$. This is due to the fact that the baryon
and the constituent quarks cannot be on their mass shell simultaneously.
However, $\gamma^{+}u(\bar{P})=\gamma^{+}u(P)$ holds~\cite{Chua:2018lfa}.
The internal momenta are defined as
\begin{align}
k_{i}=(k_{i}^{-},k_{i}^{+},k_{i\perp}) & =(e_{i}-k_{iz},e_{i}+k_{iz},k_{i\perp})\nonumber \\
 & =(\frac{m_{i}^{2}+k_{i\perp}^{2}}{x_{i}M_{0}},x_{i}M_{0},k_{i\perp}),
\end{align}
then it is easy to obtain 
\begin{align}
e_{i} & =\frac{x_{i}M_{0}}{2}+\frac{m_{i}^{2}+k_{i\perp}^{2}}{2x_{i}M_{0}},\nonumber \\
k_{iz} & =\frac{x_{i}M_{0}}{2}-\frac{m_{i}^{2}+k_{i\perp}^{2}}{2x_{i}M_{0}},
\end{align}
where $e_{i}$ denotes the energy of the $i$-th quark in the rest
frame of $\bar{P}$. The momenta $k_{i\perp}$ and $k_{iz}$ constitute
a momentum 3-vector $\vec{k}_{i}=(k_{i\perp},k_{iz})$.

For $\Lambda_{Q}$, in which the $u,d$ quarks are considered
to form a $0^{+}$ diquark, $\Psi$ in Eq. (\ref{eq:baryon_state})
can be shown as
\begin{align}
 & \Psi_{0}^{S=\frac{1}{2},S_{z}}(\tilde{p}_{i},\lambda_{i})\nonumber \\
= & A_{0}\bar{u}(p_{3},\lambda_{3})(\bar{\slashed P}+M_{0})(-\gamma_{5})C\bar{u}^{T}(p_{2},\lambda_{2})\nonumber \\
 & \times\bar{u}(p_{1},\lambda_{1})u(\bar{P},S_{z})\Phi(x_{i},k_{i\perp}),\label{eq:wf_LQ}
\end{align}
for $\Sigma_{Q}$, in which the $u,d$ quarks are considered
to form a $1^{+}$ diquark,
\begin{align}
 & \Psi_{1}^{S=\frac{1}{2},S_{z}}(\tilde{p}_{i},\lambda_{i})\nonumber \\
= & A_{1}\bar{u}(p_{3},\lambda_{3})(\bar{\slashed P}+M_{0})(\gamma^{\mu}-v^{\mu})C\bar{u}^{T}(p_{2},\lambda_{2})\nonumber \\
 & \times\bar{u}(p_{1},\lambda_{1})(\frac{1}{\sqrt{3}}\gamma_{\mu}\gamma_{5})u(\bar{P},S_{z})\Phi(x_{i},k_{i\perp}),\label{eq:wf_SQ}
\end{align}
and for $\Sigma_{Q}^{*}$, in which the $u,d$ quarks are also
considered to form a $1^{+}$ diquark,
\begin{align}
 & \Psi_{1}^{S=\frac{3}{2},S_{z}}(\tilde{p}_{i},\lambda_{i})\nonumber \\
= & A_{1}^{\prime}\bar{u}(p_{3},\lambda_{3})(\bar{\slashed P}+M_{0})(\gamma^{\mu}-v^{\mu})C\bar{u}^{T}(p_{2},\lambda_{2})\nonumber \\
 & \times\bar{u}(p_{1},\lambda_{1})u_{\mu}(\bar{P},S_{z})\Phi(x_{i},k_{i\perp}),\label{eq:wf_SQstar}
\end{align}
where $v^{\mu}\equiv\bar{P}^{\mu}/M_{0}$, and $\Phi$ is the momentum wavefunction. 
The proofs of Eqs.~(\ref{eq:wf_LQ}-\ref{eq:wf_SQstar})
can be found in Appendix~\ref{app:wave_functions}.

With the normalization of the baryon state 
\begin{align}
\langle{\cal B}(P^{\prime},S^{\prime},S_{z}^{\prime})|{\cal B}(P,S,S_{z})\rangle= & 2(2\pi)^{3}P^{+}\delta^{3}(\tilde{P}^{\prime}-\tilde{P})\nonumber \\
 & \times\delta_{S^{\prime}S}\delta_{S_{z}^{\prime}S_{z}},\label{eq:state_normalization}
\end{align}
and
\begin{align}
\int\left(\prod_{i=1}^{3}\frac{dx_{i}d^{2}k_{i\perp}}{2(2\pi)^{3}}\right)2(2\pi)^{3}\delta(1-\sum x_{i})\delta^{2}(\sum k_{i\perp})\nonumber \\
\times|\Phi(x_{i},k_{i\perp})|^{2}=1,
\end{align}
one can obtain 
\begin{equation}
A_{0}=A_{1}=A_{1}^{\prime}=\frac{1}{4\sqrt{M_{0}^{3}(e_{1}+m_{1})(e_{2}+m_{2})(e_{3}+m_{3})}}.\label{eq:normalization_factor}
\end{equation}

The momentum wavefunction can be given by 
\begin{equation}
\Phi(x_{i},k_{i\perp})=\sqrt{\frac{e_{1}e_{2}e_{3}}{x_{1}x_{2}x_{3}M_{0}}}\varphi(\vec{k}_{1},\beta_{1})\varphi(\frac{\vec{k}_{2}-\vec{k}_{3}}{2},\beta_{23})\label{eq:momentum_wf}
\end{equation}
where $\varphi(\vec{k},\beta)\equiv4\left(\frac{\pi}{\beta^{2}}\right)^{3/4}\exp\left(\frac{-k_{\perp}^{2}-k_{z}^{2}}{2\beta^{2}}\right)$,
and $\beta_{1}$ and $\beta_{23}$ are the shape parameters.

Some important notes are given below.
\begin{itemize}
\item 
The definition of a baryon state is the most important part of LFQM, while the spin wavefunction is the most important part in the definition of a baryon state.
It is worth pointing out that, when we arrive at Eqs.~(\ref{eq:wf_LQ}-\ref{eq:wf_SQstar}), we do not introduce any additional assumptions, for example,
it does not assume heavy quark symmetry,
nor does it depend on the coordinate system selection of LFQM (see below).
Therefore, the spin wavefunctions in Eqs.~(\ref{eq:wf_LQ}-\ref{eq:wf_SQstar}) may not only apply to heavy flavor baryons, but should also apply to light flavor baryons. 
\item 
From the proof in Appendix~\ref{app:wave_functions}, one can clearly see that,
the Lorentz boost between the rest frame of ``diquark" and the rest frame of $\bar{P}$
plays a crucial role. The proof of Eq.~(\ref{eq:wf_LQ}) is relatively simple, this is because the first case involves a scalar diquark, whose Lorentz boost is trivial. The proofs of Eqs.~(\ref{eq:wf_SQ}) and~(\ref{eq:wf_SQstar}) are relatively complicated,  because the latter two cases involve an axis-vector diquark, whose Lorentz boost is non-trivial.

\item If we only consider the spin coupling, in principle, we can choose
any two quarks for spin coupling first. However, when we consider
the flavor wavefunction, the two quarks that are coupled first are usually already
determined. For example, for $\Lambda_{Q}$, whose flavor wavefunction is $(ud-du)Q/\sqrt{2}$,
we couple the $u$ and $d$ quarks first; while for $\Xi_{cc}^{++}$,
whose flavor wavefunction is just $ccu$, we couple the two charm quarks first.
\item If identical quarks are contained in the baryon state, some additional
factor should be added. For example, for $\Xi_{cc}^{++}$, when we
calculate $\langle{\cal B}(P^{\prime},S^{\prime},S_{z}^{\prime})|{\cal B}(P,S,S_{z})\rangle$
to normalize the baryon state, a factor 2 appears because of two equivalent
contractions. An additional factor $1/\sqrt{2}$ should be added in
the definition of $|\Xi_{cc}^{++}\rangle$ in order to keep Eq.~(\ref{eq:state_normalization}) unchanged.
\end{itemize}

\subsection{To determine the shape parameters}

The shape parameters in Eq.~(\ref{eq:momentum_wf}) characterize the momentum
distribution inside the baryon. In this work, we suggest that the shape parameters
can be determined by the pole residue of the baryon, whose numerical
result can be taken from, for example, Lattice QCD or QCD sum rules.

Taking $\Lambda_{Q}$ as an example, let us focus on the matrix element
$\langle0|J_{\Lambda_{Q}}|\Lambda_{Q}\rangle$ with $J_{\Lambda_{Q}}=\epsilon_{abc}[u_{a}^{T}C\gamma_{5}d_{b}]Q_{c}$.
On the one hand, this matrix element can be calculated in LFQM
\begin{align}
 & \langle0|J_{\Lambda_{Q}}|\Lambda_{Q}(P,S_{z})\rangle\nonumber \\
 & =\int\frac{dx_{2}d^{2}k_{2\perp}}{2(2\pi)^{3}}\frac{dx_{3}d^{2}k_{3\perp}}{2(2\pi)^{3}}\frac{1}{\sqrt{x_{1}x_{2}x_{3}}}\Phi(x_{i},k_{i\perp})\sqrt{6}A_{0}\nonumber \\
 & \times{\rm Tr}[C\gamma_{5}(\slashed p_{3}+m_{3})(\bar{\slashed P}+M_{0})(-\gamma_{5})C(\slashed p_{2}+m_{2})^{T}]\nonumber \\
 & \times(\slashed p_{1}+m_{1})u(\bar{P},S_{z}).\label{eq:residue_in_LFQM}
\end{align}
On the other hand, the pole residue of baryon is defined by
\begin{equation}
\langle0|J_{\Lambda_{Q}}|\Lambda_{Q}(P,S_{z})\rangle=\lambda_{\Lambda_{Q}}u(P,S_{z}).\label{eq:residue_definition}
\end{equation}
Respectively multiplying Eqs.~(\ref{eq:residue_in_LFQM}) and (\ref{eq:residue_definition})
with $\sum_{S_{z}}\bar{u}(P,S_{z})\gamma^{+}$ from the left, also
noting that $\gamma^{+}u(P)=\gamma^{+}u(\bar{P})$, one can arrive at
\begin{align}
 & \int\frac{dx_{2}d^{2}k_{2\perp}}{2(2\pi)^{3}}\frac{dx_{3}d^{2}k_{3\perp}}{2(2\pi)^{3}}\frac{1}{\sqrt{x_{1}x_{2}x_{3}}}\Phi(x_{i},k_{i\perp})\sqrt{6}A_{0}\nonumber \\
 & \times{\rm Tr}[C\gamma_{5}(\slashed p_{3}+m_{3})(\bar{\slashed P}+M_{0})(-\gamma_{5})C(\slashed p_{2}+m_{2})^{T}]\nonumber \\
 & \times{\rm Tr}[\gamma^{+}(\slashed p_{1}+m_{1})(\bar{\slashed P}+M_{0})],\label{eq:residue_left_multiply}
\end{align}
and
\begin{equation}
\lambda_{\Lambda_{Q}}{\rm Tr}[\gamma^{+}(\slashed P+M)].\label{eq:residue_def_left_multiply}
\end{equation}
Equating Eqs.~(\ref{eq:residue_left_multiply})
and (\ref{eq:residue_def_left_multiply}), one can obtain the expression
of pole residue in LFQM 
\begin{align}
\lambda_{\Lambda_{Q}}= & \frac{1}{{\rm Tr}[\gamma^{+}(\slashed P+M)]}\nonumber \\
 & \times\int\frac{dx_{2}d^{2}k_{2\perp}}{2(2\pi)^{3}}\frac{dx_{3}d^{2}k_{3\perp}}{2(2\pi)^{3}}\frac{1}{\sqrt{x_{1}x_{2}x_{3}}}\Phi(x_{i},k_{i\perp})\sqrt{6}A_{0}\nonumber \\
 & \times{\rm Tr}[C\gamma_{5}(\slashed p_{3}+m_{3})(\bar{\slashed P}+M_{0})(-\gamma_{5})C(\slashed p_{2}+m_{2})^{T}]\nonumber \\
 & \times{\rm Tr}[\gamma^{+}(\slashed p_{1}+m_{1})(\bar{\slashed P}+M_{0})],\label{eq:residue_form_1}
\end{align}
which can be used to determine the shape parameters in $\Phi$ 
provided the pole residue is known. Since there are two shape parameters
in $\Phi$, one more equation is desirable, at this time, use $\sum_{S_{z}}\bar{u}(P,S_{z})\gamma^{+}\gamma^{-}$
to left multiply instead, finally we have 
\begin{align}
\lambda_{\Lambda_{Q}}= & \frac{1}{{\rm Tr}[\gamma^{+}\gamma^{-}(\slashed P+M)]}\nonumber \\
 & \times\int\frac{dx_{2}d^{2}k_{2\perp}}{2(2\pi)^{3}}\frac{dx_{3}d^{2}k_{3\perp}}{2(2\pi)^{3}}\frac{1}{\sqrt{x_{1}x_{2}x_{3}}}\Phi(x_{i},k_{i\perp})\sqrt{6}A_{0}\nonumber \\
 & \times{\rm Tr}[C\gamma_{5}(\slashed p_{3}+m_{3})(\bar{\slashed P}+M_{0})(-\gamma_{5})C(\slashed p_{2}+m_{2})^{T}]\nonumber \\
 & \times{\rm Tr}[\gamma^{+}\gamma^{-}(\slashed p_{1}+m_{1})(\bar{\slashed P}+M_{0})].\label{eq:residue_form_2}
\end{align}

The expressions of pole residues of $\Sigma_{Q}$ and $\Xi_{cc}$
can also be obtained in a similar way.

In addition, it should be noted that, the baryon mass $M$ can in
turn be extracted by equating Eqs.~(\ref{eq:residue_form_1}) and
(\ref{eq:residue_form_2}) once the shape parameters are fixed by, for example, global fitting. 

\subsection{Form factors of $\Lambda_{b}\to\Lambda_{c}$ }

\label{subsec:Lb2Lc}

On the one hand, the weak decay matrix element $\langle\Lambda_{c}|\bar{c}\gamma^{\mu}(1-\gamma_{5})b|\Lambda_{b}\rangle$
can be parameterized in terms of form factors
\begin{align}
 & \langle\Lambda_{c}(P^{\prime},S_{z}^{\prime})|\bar{c}\gamma^{\mu}(1-\gamma_{5})b|\Lambda_{b}(P,S_{z})\rangle\nonumber \\
= & \bar{u}(P^{\prime},S_{z}^{\prime})\Big\{\Big[\gamma^{\mu}f_{1}(q^{2})+i\sigma^{\mu\nu}\frac{q_{\nu}}{M}f_{2}(q^{2})+\frac{q^{\mu}}{M}f_{3}(q^{2})\Big]\nonumber \\
 & -\Big[\gamma^{\mu}g_{1}(q^{2})+i\sigma^{\mu\nu}\frac{q_{\nu}}{M}g_{2}(q^{2})+\frac{q^{\mu}}{M}g_{3}(q^{2})\Big]\gamma_{5}\Big\} u(P,S_{z}),\label{eq:ff_parameterization}
\end{align}
where $q=P-P^{\prime}$, and $f_{i}$, $g_{i}$ are the form factors. 
On the other hand, the matrix element can also be calculated in LFQM
\begin{align}
 & \langle\Lambda_{c}(P^{\prime},S_{z}^{\prime})|\bar{c}\gamma^{\mu}(1-\gamma_{5})b|\Lambda_{b}(P,S_{z})\rangle\nonumber \\
= & \int\{d^{3}\tilde{p}_{2}\}\{d^{3}\tilde{p}_{3}\}\frac{A_{0}^{\prime}A_{0}}{\sqrt{p_{1}^{\prime+}p_{1}^{+}P^{\prime+}P^{+}}}\Phi^{\prime*}(x_{i}^{\prime},k_{i\perp}^{\prime})\Phi(x_{i},k_{i\perp})\nonumber \\
 & \times{\rm Tr}[(\bar{\slashed P}+M_{0})(-\gamma_{5})C(\slashed p_{2}+m_{2})^{T}C\gamma_{5}(\bar{\slashed P}^{\prime}+M_{0}^{\prime})\nonumber \\
 & \qquad(\slashed p_{3}+m_{3})]\nonumber \\
 & \times\bar{u}(\bar{P}^{\prime},S_{z}^{\prime})(\slashed p_{1}^{\prime}+m_{1}^{\prime})\gamma^{\mu}(1-\gamma_{5})(\slashed p_{1}+m_{1})u(\bar{P},S_{z}).\label{eq:matrix_element_Lb_Lc}
\end{align}

Now we extract the form factors $f_{1,2}$ and $g_{1,2}$ in the following
method. Respectively multiplying the ``+'' component of the vector current
part of Eq.~(\ref{eq:ff_parameterization}) by $\sum_{S_{z},S_{z}^{\prime}}\bar{u}(P,S_{z})\gamma^{+}u(P^{\prime},S_{z}^{\prime})$
and $\sum_{S_{z},S_{z}^{\prime}}\bar{u}(P,S_{z})(\sum_{j=1}^{2}i\sigma^{+j}q^{j})u(P^{\prime},S_{z}^{\prime})$
from the left, one can obtain 
\begin{align}
{\rm Tr}[(\slashed P+M)\gamma^{+}(\slashed P^{\prime}+M^{\prime})(f_{1}\gamma^{+}+f_{2}i\sigma^{+\nu}\frac{q_{\nu}}{M})]\nonumber \\
=8P^{+}P^{\prime+}f_{1},
\end{align}
and
\begin{align}
 & {\rm Tr}[(\slashed P+M)(\sum_{j=1}^{2}i\sigma^{+j}q^{j})(\slashed P^{\prime}+M^{\prime})\nonumber \\
 & \quad(f_{1}\gamma^{+}+f_{2}i\sigma^{+\nu}\frac{q_{\nu}}{M})]=-8P^{+}P^{\prime+}\frac{q^{2}}{M}f_{2}.
\end{align}
Respectively multiplying the ``+'' component of the axial-vector current part of Eq.~(\ref{eq:ff_parameterization}) by $\sum_{S_{z},S_{z}^{\prime}}\bar{u}(P,S_{z})\gamma^{+}\gamma_{5}u(P^{\prime},S_{z}^{\prime})$
and $\sum_{S_{z},S_{z}^{\prime}}\bar{u}(P,S_{z})(\sum_{j=1}^{2}i\sigma^{+j}q^{j}\gamma_{5})u(P^{\prime},S_{z}^{\prime})$
from the left, one can obtain 
\begin{align}
{\rm Tr}[(\slashed P+M)\gamma^{+}\gamma_{5}(\slashed P^{\prime}+M^{\prime})(g_{1}\gamma^{+}+g_{2}i\sigma^{+\nu}\frac{q_{\nu}}{M})\gamma_{5}]\nonumber \\
=8P^{+}P^{\prime+}g_{1},
\end{align}
and 
\begin{align}
 & {\rm Tr}[(\slashed P+M)(\sum_{j=1}^{2}i\sigma^{+j}q^{j}\gamma_{5})(\slashed P^{\prime}+M^{\prime})\nonumber \\
 & \quad(g_{1}\gamma^{+}+g_{2}i\sigma^{+\nu}\frac{q_{\nu}}{M})\gamma_{5}]=8P^{+}P^{\prime+}\frac{q^{2}}{M}g_{2}.
\end{align}
Then doing the same thing for Eq.~(\ref{eq:matrix_element_Lb_Lc}),
one can obtain 
\begin{align}
f_{1} = & \frac{1}{8P^{+}P^{\prime+}}\int\{d^{3}\tilde{p}_{2}\}\{d^{3}\tilde{p}_{3}\}\frac{\Phi^{\prime*}\Phi}{\sqrt{p_{1}^{\prime+}p_{1}^{+}P^{\prime+}P^{+}}}A_{0}^{\prime}A_{0}\nonumber \\
 & \times {\rm Tr}[(\bar{\slashed P}+M_{0})(-\gamma_{5})C(\slashed p_{2}+m_{2})^{T}C\gamma_{5}(\bar{\slashed P}^{\prime}+M_{0}^{\prime})\nonumber \\
 & \qquad(\slashed p_{3}+m_{3})]\nonumber \\
 & \times{\rm Tr}[(\bar{\slashed P}+M_{0})\Gamma_{1}(\bar{\slashed P}^{\prime}+M_{0}^{\prime})(\slashed p_{1}^{\prime}+m_{1}^{\prime})\Gamma_{2}(\slashed p_{1}+m_{1})]\label{eq:f1_Lb_Lc}
\end{align}
with 
\begin{equation}
\Gamma_{1}=\gamma^{+},\quad\Gamma_{2}=\gamma^{+},
\end{equation}
where we have used $\gamma^{+}u(P)=\gamma^{+}u(\bar{P})$.
Same expressions
can be obtained for $f_{2}$ and $g_{1,2}$, except that, 
\begin{itemize}
\item for $f_{2}$,
\begin{equation}
\Gamma_{1}=-\frac{M}{q^{2}}\sum_{j=1}^{2}i\sigma^{+j}q^{j},\quad\Gamma_{2}=\gamma^{+},
\end{equation}
\item for $g_{1}$,
\begin{equation}
\Gamma_{1}=\gamma^{+}\gamma_{5},\quad\Gamma_{2}=\gamma^{+}\gamma_{5},
\end{equation}
\item for $g_{2}$,
\begin{equation}
\Gamma_{1}=\frac{M}{q^{2}}\sum_{j=1}^{2}i\sigma^{+j}q^{j}\gamma_{5},\quad\Gamma_{2}=\gamma^{+}\gamma_{5}.
\end{equation}
\end{itemize}

In practice, we choose the frame that satisfies $q^{+}=0$, that is,
\begin{equation}
P^{+}-P^{\prime+}=0.
\end{equation}
When calculating the weak decay matrix element in Eq.~(\ref{eq:matrix_element_Lb_Lc}), one can find that
the momenta of quark 2 and quark 3 remain unchanged from the initial state to the final state,
from which one can obtain
\begin{align}
 & x_{2}^{\prime}=x_{2},\quad k_{2\perp}^{\prime}=k_{2\perp}+x_{2}q_{\perp},\nonumber \\
 & x_{3}^{\prime}=x_{3},\quad k_{3\perp}^{\prime}=k_{3\perp}+x_{3}q_{\perp},
\end{align}
and furthermore,
\begin{equation}
x_{1}^{\prime}=x_{1},\quad k_{1\perp}^{\prime}=k_{1\perp}-(1-x_{1})q_{\perp}.
\end{equation}

One comment. As pointed out in Ref.~\cite{Ke:2019smy}, the form factors
$f_{3}$ and $g_{3}$ cannot be extracted for we have imposed the
condition $q^{+}=0$. However, these two form factors do not contribute
to the $1/2\to1/2$ semileptonic decays if the electron mass
is neglected.

\subsection{Form factors of $\Sigma_{b}\to\Sigma_{c}$ }

$\langle\Sigma_{c}|\bar{c}\gamma^{\mu}(1-\gamma_{5})b|\Sigma_{b}\rangle$
can also be otained in LFQM as 
\begin{align}
 & \langle\Sigma_{c}(P^{\prime},S_{z}^{\prime})|\bar{c}\gamma^{\mu}(1-\gamma_{5})b|\Sigma_{b}(P,S_{z})\rangle\nonumber \\
= & \int\{d^{3}\tilde{p}_{2}\}\{d^{3}\tilde{p}_{3}\}\frac{A_{1}^{\prime}A_{1}}{\sqrt{p_{1}^{\prime+}p_{1}^{+}P^{\prime+}P^{+}}}\Phi^{\prime*}(x_{i}^{\prime},k_{i\perp}^{\prime})\Phi(x_{i},k_{i\perp})\nonumber \\
 &\times {\rm Tr}[(\bar{\slashed P}+M_{0})(\gamma^{\rho}-v^{\rho})C(\slashed p_{2}+m_{2})^{T}C(\gamma^{\sigma}-v^{\prime\sigma})\nonumber \\
 & \qquad(\bar{\slashed P}^{\prime}+M_{0}^{\prime})(\slashed p_{3}+m_{3})]\nonumber \\
 & \times\bar{u}(\bar{P}^{\prime},S_{z}^{\prime})\frac{1}{\sqrt{3}}\gamma_{\sigma}\gamma_{5}(\slashed p_{1}^{\prime}+m_{1}^{\prime})\gamma^{\mu}(1-\gamma_{5})(\slashed p_{1}+m_{1})\nonumber \\
 & \quad\frac{1}{\sqrt{3}}\gamma_{\rho}\gamma_{5}u(\bar{P},S_{z}),\label{eq:matrix_element_Sb_Sc}
\end{align}
where $v^{\mu}=\bar{P}^{\mu}/M_{0}$ and $v^{\prime\mu}=\bar{P}^{\prime\mu}/M_{0}^{\prime}$.
It turns out that 
\begin{align}
f_{1}= & \frac{1}{8P^{+}P^{\prime+}}\int\{d^{3}\tilde{p}_{2}\}\{d^{3}\tilde{p}_{3}\}\frac{\Phi^{\prime*}\Phi}{\sqrt{p_{1}^{\prime+}p_{1}^{+}P^{\prime+}P^{+}}}A_{1}^{\prime}A_{1}\nonumber \\
 & \times{\rm Tr}[(\bar{\slashed P}+M_{0})(\gamma^{\rho}-v^{\rho})C(\slashed p_{2}+m_{2})^{T}C(\gamma^{\sigma}-v^{\prime\sigma})\nonumber \\
 & \qquad(\bar{\slashed P}^{\prime}+M_{0}^{\prime})(\slashed p_{3}+m_{3})]\nonumber \\
 & \times{\rm Tr}[(\bar{\slashed P}+M_{0})\Gamma_{1}(\bar{\slashed P}^{\prime}+M_{0}^{\prime})\frac{1}{\sqrt{3}}\gamma_{\sigma}\gamma_{5}(\slashed p_{1}^{\prime}+m_{1}^{\prime})\Gamma_{2}\nonumber \\
 & \qquad(\slashed p_{1}+m_{1})\frac{1}{\sqrt{3}}\gamma_{\rho}\gamma_{5}]\label{eq:f1_Sb_Sc}
\end{align}
with 
\begin{equation}
\Gamma_{1}=\gamma^{+},\quad\Gamma_{2}=\gamma^{+}.
\end{equation}
$f_{2}$ and $g_{1,2}$ can also be obtained by the same assignments to $\Gamma_{1,2}$
as those in Subsec.~\ref{subsec:Lb2Lc}. 

\subsection{Form factors of $\Xi_{cc}\to\Lambda_{c}$ }

\label{subsec:Xicc2Lc}

$\langle\Lambda_{c}|\bar{d}\gamma^{\mu}(1-\gamma_{5})c|\Xi_{cc}\rangle$
can also be obtained in LFQM as 
\begin{align}
 & \langle\Lambda_{c}(P^{\prime},S_{z}^{\prime})|\bar{d}\gamma^{\mu}(1-\gamma_{5})c|\Xi_{cc}(P,S_{z})\rangle\nonumber \\
= & \int\{d^{3}\tilde{p}_{2}\}\{d^{3}\tilde{p}_{3}\}\frac{A_{0}^{\prime}A_{1}}{\sqrt{p_{1}^{\prime+}p_{1}^{+}P^{\prime+}P^{+}}}\Phi^{\prime*}(x_{i}^{\prime},k_{i\perp}^{\prime})\Phi(x_{i},k_{i\perp})\nonumber \\
 & \times\frac{2}{\sqrt{2}}\bar{u}(\bar{P}^{\prime},S_{z}^{\prime})(\slashed p_{2}+m_{2})(\bar{\slashed P}+M_{0})(\gamma^{\nu}-v^{\nu})C\nonumber\\
 & \quad(\slashed p_{1}+m_{1})^{T}(1-\gamma_{5})^{T}\gamma^{\mu T}(\slashed p_{1}^{\prime}+m_{1}^{\prime})^{T}C\gamma_{5}\nonumber \\
 & \quad(\bar{\slashed P}^{\prime}+M_{0}^{\prime})(\slashed p_{3}+m_{3})\frac{1}{\sqrt{3}}\gamma_{\nu}\gamma_{5}u(\bar{P},S_{z}),\label{eq:matrix_element_Xicc_Lc}
\end{align}
where the factor 2 comes from the two equivalent contractions, and
the factor $1/\sqrt{2}$ comes from the normalization of $\Xi_{cc}^{++}$ state,
which has been pointed out in Subsec.~\ref{subsec:The-baryon-states}. It turns out that 
\begin{align}
f_{1} & =\frac{1}{8P^{+}P^{\prime+}}\int\{d^{3}\tilde{p}_{2}\}\{d^{3}\tilde{p}_{3}\}\frac{\Phi^{\prime*}\Phi}{\sqrt{p_{1}^{\prime+}p_{1}^{+}P^{\prime+}P^{+}}}A_{0}^{\prime}A_{1}\nonumber \\
 & \times\frac{2}{\sqrt{2}}{\rm Tr}[(\bar{\slashed P}+M_{0})\Gamma_{1}(\bar{\slashed P}^{\prime}+M_{0}^{\prime})(\slashed p_{2}+m_{2})(\bar{\slashed P}+M_{0})\nonumber \\
 & \quad(\gamma^{\nu}-v^{\nu})(\slashed p_{1}-m_{1})\Gamma_{2}(\slashed p_{1}^{\prime}-m_{1}^{\prime})\gamma_{5}(\bar{\slashed P}^{\prime}+M_{0}^{\prime})\nonumber \\
 & \quad(\slashed p_{3}+m_{3})\frac{1}{\sqrt{3}}\gamma_{\nu}\gamma_{5}]\label{eq:f1_Xicc2Lc}
\end{align}
with 
\begin{equation}
\Gamma_{1}=\gamma^{+},\quad\Gamma_{2}=\gamma^{+}.
\end{equation}
$f_{2}$ and $g_{1,2}$ can also be obtained by the same assignments
to $\Gamma_{1,2}$ as those in Subsec.~\ref{subsec:Lb2Lc} except for
the only one difference for $g_{1,2}$ 
\begin{align}
\Gamma_{2}= & \gamma_{5}\gamma^{+},
\end{align}
because we have performed a transpose in Eq.~(\ref{eq:matrix_element_Xicc_Lc}).

\subsection{The relation between the two pictures}

Define the spin wavefunction in Eq.~(\ref{eq:wf_LQ}) as $\psi_{0}(321)$ and
that in Eq.~(\ref{eq:wf_SQ}) as $\psi_{1}(321)$, i.e.,
\begin{align}
\psi_{0}(321) & \equiv\bar{u}(p_{3},\lambda_{3})(\bar{\slashed P}+M_{0})(-\gamma_{5})C\bar{u}^{T}(p_{2},\lambda_{2})\nonumber \\
 & \times\bar{u}(p_{1},\lambda_{1})u(\bar{P},S_{z}),\nonumber \\
\psi_{1}(321) & \equiv\bar{u}(p_{3},\lambda_{3})(\bar{\slashed P}+M_{0})(\gamma^{\mu}-v^{\mu})C\bar{u}^{T}(p_{2},\lambda_{2})\nonumber \\
 & \times\bar{u}(p_{1},\lambda_{1})(\frac{1}{\sqrt{3}}\gamma_{\mu}\gamma_{5})u(\bar{P},S_{z}).\label{eq:spin_wfs}
\end{align}
$\psi_{0,1}(321)$ have the same normalization factor as that in Eq.~(\ref{eq:normalization_factor}). Moreover, $\psi_{0,1}(321)$ are
orthogonal
\begin{equation}
\sum_{\lambda_{1}\lambda_{2}\lambda_{3}}\psi_{0}^{\dagger}(321^{\prime})\psi_{1}(321)=0,
\end{equation}
where quark $1^{\prime}$ does not need to be the same as quark 1. 

Now consider the relation between the three-quark picture and
the diquark picture. The key observation is that in the three-quark
picture, the first two quarks form a diquark in the diquark picture.
Specifically, in $\psi_{0}(321)/\psi_{1}(321)$, quark 3 and quark 2 
are considered to form a scalar/axial-vector diquark. In fact,
$\psi_{0,1}(321)$ constitute a diquark basis. 

One can easily check that
\begin{align}
\psi_{0}(321) & =-\psi_{0}(231),\nonumber \\
\psi_{1}(321) & =\psi_{1}(231).
\end{align}
In addition, it can be shown that
\begin{equation}
\left(\begin{array}{c}
\psi_{0}(312)\\
\psi_{1}(312)
\end{array}\right)=T\left(\begin{array}{c}
\psi_{0}(321)\\
\psi_{1}(321)
\end{array}\right),
\end{equation}
with the transition matrix
\begin{equation}
T=\left(\begin{array}{cc}
\frac{1}{2} & -\frac{\sqrt{3}}{2}\\
-\frac{\sqrt{3}}{2} & -\frac{1}{2}
\end{array}\right),
\end{equation}
which satisfies $T^{-1}=T$, as expected.

Now we are ready to determine the overlap factors in the diquark picture.
Take the process of $\Xi_{bc}^{+}(cbu)\to\Lambda_{b}(dbu)$ as an
example, where the $b,c$ quarks in the initial state are considered
to form an axial-vector diquark, while the $u,d$ quarks in
the final state are considered to form a scalar diquark. The initial
and final states can be rewritten as
\begin{align}
\psi_{1}(bcu) & =-\frac{\sqrt{3}}{2}\psi_{0}(buc)-\frac{1}{2}\psi_{1}(buc),\nonumber \\
\psi_{0}(udb) & =\frac{1}{2}\psi_{0}(ubd)-\frac{\sqrt{3}}{2}\psi_{1}(ubd)\nonumber \\
 & =-\frac{1}{2}\psi_{0}(bud)-\frac{\sqrt{3}}{2}\psi_{1}(bud).
\end{align}
Then it is easy to write the transition matrix element as 
\begin{align}
\langle\psi_{0}(udb)|\psi_{1}(bcu)\rangle & =\frac{\sqrt{3}}{4}\langle\psi_{0}(bud)|\psi_{0}(buc)\rangle\nonumber \\
 & +\frac{\sqrt{3}}{4}\langle\psi_{1}(bud)|\psi_{1}(buc)\rangle,
\end{align}
from which, one can read the two overlap factors
\begin{equation}
c_{S}=c_{A}=\frac{\sqrt{3}}{4}.
\end{equation}
They are the same as those in the diquark picture in Ref.~\cite{Wang:2017mqp}.

It's time to go one step further to derive the overlap factors of
$\Xi_{cc}^{++}(ccu)\to\Lambda_{c}(dcu)$. To this end, notice that
there is only one difference between this process and $\Xi_{bc}^{+}(cbu)\to\Lambda_{b}(dbu)$,
that is, $\Xi_{cc}^{++}$ contains two identical quarks. At this time,
one can obtain the overlap factors for $\Xi_{cc}^{++}\to\Lambda_{c}$
\begin{equation}
c_{S}=c_{A}=\frac{2}{\sqrt{2}}\times\frac{\sqrt{3}}{4}=\frac{\sqrt{6}}{4},
\end{equation}
where the factor $2/\sqrt{2}$ can be found in Eq.~(\ref{eq:matrix_element_Xicc_Lc}).
These factors are also the same as those in the diquark picture in
Ref.~\cite{Wang:2017mqp}.

\subsection{Improved definitions of interpolating currents}

The definitions of interpolating currents are the starting point of
Lattice QCD and QCD sum rules. The following definitions are
usually adopted for $\Lambda_{Q}$ and $\Sigma_{Q}$ in the literature~\cite{Colangelo:2000dp,Wang:2010fq,Wang:2009cr,Wang:2010hs}
\begin{align}
J_{\Lambda_{Q}}= & \epsilon_{abc}[u_{a}^{T}C\gamma_{5}d_{b}]Q_{c},\nonumber \\
J_{\Sigma_{Q}}= & \epsilon_{abc}[u_{a}^{T}C\gamma^{\mu}d_{b}]\gamma_{\mu}\gamma_{5}Q_{c}.\label{eq:traditional_defs}
\end{align}
These interpolating currents of baryons
were first given in Ref.~\cite{Ioffe:1981kw}, and then widely used
to study the properties of baryons. 
They are obtained with the help of symmetry analysis,
however, to our knowledge, there is no literature that provides a
rigorous proof starting from quark spinors and Dirac matrices. This work
may fill this gap. 

Hermite conjugating Eq.~(\ref{eq:spin_wfs}),
one can obtain the following improved definitions of interpolating
currents for $\Lambda_{Q}$ and $\Sigma_{Q}$ 
\begin{align}
J_{\Lambda_{Q}}= & \epsilon_{abc}[u_{a}^{T}C\gamma_{5}(1+\slashed v)d_{b}]Q_{c},\nonumber \\
J_{\Sigma_{Q}}= & \epsilon_{abc}[u_{a}^{T}C(\gamma^{\mu}-v^{\mu})(1+\slashed v)d_{b}]\frac{1}{\sqrt{3}}\gamma_{\mu}\gamma_{5}Q_{c},\label{eq:better_defs}
\end{align}
where $v^{\mu}\equiv p^{\mu}/\sqrt{p^{2}}$ with $p$ the four momentum
of baryon. 
Some comments are in order. 
\begin{itemize}
\item As can be seen in Eq.~(\ref{eq:pole_residues_QCDSR}) that, $\lambda_{\Sigma_{Q}}\approx2\ \lambda_{\Lambda_{Q}}$,
which are calculated in QCD sum rules using the definitions in Eq.~(\ref{eq:traditional_defs})~\cite{Wang:2010fq,Wang:2009cr}. If the factor $1/\sqrt{3}$
in Eq.~(\ref{eq:better_defs}) is considered, one would expect $\lambda_{\Sigma_{Q}}\approx\lambda_{\Lambda_{Q}}$. 
\item It can be seen that, we can even let $v\to0$ in Eq.~(\ref{eq:better_defs})
to get the definitions in Eq.~(\ref{eq:traditional_defs}) if we temporarily forget 
the coefficient $1/\sqrt{3}$. However, we cannot
do that, because $v^{\mu}$, according to its definition, is ${\cal O}(1)$.
\end{itemize}

\section{Numerical results}

The following quark masses are adopted:
\begin{align}
 & m_{u}=m_{d}=0.25\ {\rm GeV},\nonumber \\
 & m_{c}=1.4\ {\rm GeV},\quad m_{b}=4.8\ {\rm GeV},\label{eq:mass_quark}
\end{align}
which are widely used in previous literature of LFQM~\cite{Wang:2008xt,Wang:2009mi,Shi:2016gqt,Wang:2017mqp,Zhao:2018zcb,Zhao:2018mrg,Xing:2018lre,Zhao:2022vfr}.
Throughout this article, the following baryon masses are used~\cite{ParticleDataGroup:2022pth}
\begin{align}
 & m_{\Lambda_{b}}=5.620\ {\rm GeV},\quad m_{\Lambda_{c}}=2.286\ {\rm GeV},\nonumber \\
 & m_{\Sigma_{b}}=5.811\ {\rm GeV},\quad m_{\Sigma_{c}}=2.453\ {\rm GeV},\nonumber \\
 & m_{\Xi_{cc}^{++}}=3.622\ {\rm GeV}.
\end{align}

\subsection{The shape parameters}

\label{subsec:The-shape-parameters}

Using the pole residues of $\Lambda_{Q}$, $\Sigma_{Q}$ and $\Xi_{cc}$
obtained in Refs.~\cite{Wang:2010fq,Wang:2009cr,Wang:2010hs}
\begin{align}
 & \lambda_{\Lambda_{b}}=0.030\pm0.009,\quad\lambda_{\Lambda_{c}}=0.022\pm0.008,\nonumber \\
 & \lambda_{\Sigma_{b}}=0.062\pm0.018,\quad\lambda_{\Sigma_{c}}=0.045\pm0.015,\nonumber \\
 & \lambda_{\Xi_{cc}}=0.115\pm0.027,\label{eq:pole_residues_QCDSR}
\end{align}
which are calculated in QCD sum rules, one can obtain the following optimal shape
parameters  
\begin{align}
\beta_{b,[ud]} & =0.63\pm0.05\ {\rm GeV},\quad\beta_{[ud]}=0.27\pm0.03\ {\rm GeV},\nonumber \\
\beta_{c,[ud]} & =0.45\pm0.05\ {\rm GeV};\nonumber \\
\beta_{b,\{ud\}} & =0.66\pm0.04\ {\rm GeV},\quad\beta_{\{ud\}}=0.28\pm0.03\ {\rm GeV},\nonumber \\
\beta_{c,\{ud\}} & =0.49\pm0.04\ {\rm GeV};\nonumber \\
\beta_{u,\{cc\}} & =0.490\pm0.040\ {\rm GeV},\nonumber \\
\beta_{\{cc\}} & =0.400\pm0.025\ {\rm GeV}.\label{eq:beta_values}
\end{align}

Some comments are in order.
\begin{itemize}
\item Respectively denote the pole residues calculated in Eqs.~(\ref{eq:residue_form_1})
and (\ref{eq:residue_form_2}) as $\lambda_{1}$ and $\lambda_{2}$,
the pole residue in Eq.~(\ref{eq:pole_residues_QCDSR})
as $\lambda_{{\rm QCDSR}}$. By letting $\lambda_{1}\approx\lambda_{2}\approx\lambda_{{\rm QCDSR}}$,
one can determine the shape parameters in Eq.~(\ref{eq:beta_values}),
and the uncertainty comes from that of the pole residue. 
\item It would be interesting to compare the shape parameters with those
used in the diquark picture~\cite{Zhao:2018zcb}, and the latter are
in fact the shape parameters of mesons. For example, numerically, our
$\beta_{b,[ud]}$, $\beta_{c,[ud]}$ and $\beta_{[ud]}$ are respectively
close to (in units of GeV) $\beta_{b\bar{s}}=0.623$, $\beta_{c\bar{s}}=0.535$
and $\beta_{d\bar{s}}=0.393$ in Ref.~\cite{Zhao:2018zcb}. A significant
difference is found between $\beta_{\{cc\}}=0.400\ {\rm GeV}$ in this work and $\beta_{c\bar{c}}=0.753\ {\rm GeV}$ in Ref.~\cite{Zhao:2018zcb}.
It seems that the charm quark and anti-charm quark in $\eta_{c}$
are more energetic than the two charm quarks in $\Xi_{cc}$. It is worth noting that, when
deriving the pole residue expressions for $\Xi_{cc}$, a factor of
$2/\sqrt{2}$ also appears. Using the pole residue $\lambda_{\Xi_{cc}}$
in Eq.~(\ref{eq:pole_residues_QCDSR}), and having considered this
factor, one can obtain the much smaller shape parameter $\beta_{\{cc\}}$
together with $\beta_{u,\{cc\}}$. 
\end{itemize}

\subsection{The form factors and semileptonic decays }

The following form factors at $q^{2}=0$ are obtained for $\Lambda_{b}\to\Lambda_{c}$:
\begin{align}
 & f_{1}(0)=0.469\pm0.029,\quad f_{2}(0)=-0.105\pm0.011,\nonumber \\
 & g_{1}(0)=0.461\pm0.027,\quad g_{2}(0)=0.006\pm0.005,\label{eq:Lb2Lc_f0}
\end{align}
for $\Sigma_{b}\to\Sigma_{c}$:
\begin{align}
 & f_{1}(0)=0.490\pm0.018,\quad f_{2}(0)=0.467\pm0.006,\nonumber \\
 & g_{1}(0)=-0.163\pm0.005,\quad g_{2}(0)=0.007\pm0.001,\label{eq:Sb2Sc_f0}
\end{align}
and for $\Xi_{cc}\to\Lambda_{c}$:
\begin{align}
 & f_{1}(0)=0.517\pm0.071,\quad f_{2}(0)=-0.036\pm0.007,\nonumber \\
 & g_{1}(0)=0.155\pm0.019,\quad g_{2}(0)=-0.072\pm0.012.\label{eq:Xicc2Lc_f0}
\end{align}
It can be seen that about 6\%, 4\%, 14\% uncertainties are respectively
introduced for the three groups of form factors because of the uncertainty
of the pole residue.

To access the $q^{2}$ dependence of the form factors, we calculate
the form factors in an interval $q^{2}\in[-5,0]\ {\rm GeV}^{2}$ for
$\Lambda_{b}\to\Lambda_{c}$ and $\Sigma_{b}\to\Sigma_{c}$, and $q^{2}\in[-0.5,0]\ {\rm GeV}^{2}$
for $\Xi_{cc}\to\Lambda_{c}$, and fit the results with the following
simplified $z$-expansion~\cite{Detmold:2015aaa}:
\begin{equation}
f(q^{2})=\frac{a+b\ z(q^{2})}{1-q^{2}/m_{{\rm pole}}^{2}},
\end{equation}
where $m_{{\rm pole}}=m_{B_{c}}$ for $\Lambda_{b}\to\Lambda_{c}$
and $\Sigma_{b}\to\Sigma_{c}$, and $m_{{\rm pole}}=m_{D}$ for $\Xi_{cc}\to\Lambda_{c}$, 
\begin{equation}
z(q^{2})=\frac{\sqrt{t_{+}-q^{2}}-\sqrt{t_{+}-q_{{\rm max}}^{2}}}{\sqrt{t_{+}-q^{2}}+\sqrt{t_{+}-q_{{\rm max}}^{2}}}
\end{equation}
with $t_{+}=m_{{\rm pole}}^{2}$, $q_{{\rm max}}^{2}=(M_{\Lambda_{b}}-M_{\Lambda_{c}})^{2}$
for $\Lambda_{b}\to\Lambda_{c}$, $q_{{\rm max}}^{2}=(M_{\Sigma_{b}}-M_{\Sigma_{c}})^{2}$
for $\Sigma_{b}\to\Sigma_{c}$, and $q_{{\rm max}}^{2}=(M_{\Xi_{cc}}-M_{\Lambda_{c}})^{2}$
for $\Xi_{cc}\to\Lambda_{c}$. The fitted results of $(a,b)$ for
the three processes are given in Table~\ref{Tab:fit}.

The obtained form factors are then applied to semileptonic decays, it turns out that
the central values of decay widths and branching ratios are
\begin{align}
 & \Gamma(\Lambda_{b}\to\Lambda_{c}e^{-}\bar{\nu}_{e})=2.54\times10^{-14}\ {\rm GeV},\nonumber \\
 & {\cal B}(\Lambda_{b}\to\Lambda_{c}e^{-}\bar{\nu}_{e})=5.68\%,\nonumber \\
 & \Gamma(\Sigma_{b}\to\Sigma_{c}e^{-}\bar{\nu}_{e})=0.870\times10^{-14}\ {\rm GeV},\nonumber \\
 & \Gamma(\Xi_{cc}\to\Lambda_{c}e^{+}\nu_{e})=0.755\times10^{-14}\ {\rm GeV},\nonumber \\
 & {\cal B}(\Xi_{cc}\to\Lambda_{c}e^{+}\nu_{e})=0.294\%,
\end{align}
where $\tau_{\Lambda_{b}}=1.471\times10^{-12}\ {\rm s}$,
$|V_{cb}|=0.0408$, and $\tau_{\Xi_{cc}^{++}}=0.256\times10^{-12}\ {\rm s}$,
$|V_{cd}|=0.221$ have been used~\cite{ParticleDataGroup:2022pth}. Considering the uncertainties 
of form factors in Eq.~(\ref{eq:Lb2Lc_f0}-\ref{eq:Xicc2Lc_f0}),
there are respectively about 13\%, 7\%, and 29\% uncertainties in
these phenomenological predictions. 

\begin{table*}
\caption{Fitted results of $(a,b)$ for the form factors.}
\label{Tab:fit} %
\begin{tabular}{c|c|c||c|c|c||c|c|c}
\hline 
Transition & $F$ & $(a,b)$ & Transition & $F$ & $(a,b)$ & Transition & $F$ & $(a,b)$\tabularnewline
\hline 
 & $f_{1}$ & $(0.648,-2.177)$ &  & $f_{1}$ & $(0.794,-3.617)$ &  & $f_{1}$ & $(1.101,-3.312)$\tabularnewline
$\Lambda_{b}\to\Lambda_{c}$ & $f_{2}$ & $(-0.162,0.704)$ & $\Sigma_{b}\to\Sigma_{c}$ & $f_{2}$ & $(0.728,-3.103)$ & $\Xi_{cc}\to\Lambda_{c}$ & $f_{2}$ & $(-0.064,0.162)$\tabularnewline
 & $g_{1}$ & $(0.632,-2.068)$ &  & $g_{1}$ & $(-0.222,0.697)$ &  & $g_{1}$ & $(0.280,-0.710)$\tabularnewline
 & $g_{2}$ & $(0.011,-0.065)$ &  & $g_{2}$ & $(0.011,-0.053)$ &  & $g_{2}$ & $(-0.185,0.644)$\tabularnewline
\hline 
\end{tabular}
\end{table*}

\subsection{Comparison with other results in the literature}

In Table~\ref{Tab:comparision_ff} and Table~\ref{Tab:comparision_decay_widths},
we respectively compare our form factors and semileptonic decay widths
with those in the literature. It can be seen that our predictions
are comparable with other results; in addition, it seems that the
diquark picture tends to give larger predictions. It is likely that 
larger shape parameters are used in the diquark picture, as pointed
out in Subsec.~\ref{subsec:The-shape-parameters}. 

\begin{table*}
\caption{Our form factors are compared with other results in the literature.
The asterisk on Ref.~\cite{Shi:2019hbf} indicates that, in this literature,
we made a mistake in the sign for the axial-vector form factors, and
here we have corrected it. }
\label{Tab:comparision_ff} %
\begin{tabular}{c|c|c|c|c|c|c}
\hline 
$\Lambda_{b}\to\Lambda_{c}$ & $f_{1}(0)$ & $f_{2}(0)$ & $f_{3}(0)$ & $g_{1}(0)$ & $g_{2}(0)$ & $g_{3}(0)$\tabularnewline
\hline 
This work & $0.469\pm0.029$ & $-0.105\pm0.011$ & -- & $0.461\pm0.027$ & $0.006\pm0.005$ & --\tabularnewline
Three-quark~\cite{Ke:2019smy} & $0.488$ & $-0.180$ & -- & $0.470$ & $-0.048$ & --\tabularnewline
Diquark~\cite{Zhao:2018zcb} & $0.670$ & $-0.132$ & -- & $0.656$ & $-0.012$ & --\tabularnewline
Diquark~\cite{Ke:2007tg} & $0.506$ & $-0.099$ & -- & $0.501$ & $-0.009$ & --\tabularnewline
QCDSR~\cite{Zhao:2020mod} & $0.431$ & $-0.123$ & $0.022$ & $0.434$ & $0.036$ & $-0.160$\tabularnewline
LQCD~\cite{Detmold:2015aaa} & $0.418$ & $-0.099$ & $-0.075$ & $0.390$ & $-0.004$ & $-0.206$\tabularnewline
\hline 
$\Sigma_{b}\to\Sigma_{c}$ & $f_{1}(0)$ & $f_{2}(0)$ & $f_{3}(0)$ & $g_{1}(0)$ & $g_{2}(0)$ & $g_{3}(0)$\tabularnewline
\hline 
This work & $0.490\pm0.018$ & $0.467\pm0.006$ & -- & $-0.163\pm0.005$ & $0.007\pm0.001$ & --\tabularnewline
Three-quark~\cite{Ke:2019smy} & $0.494$ & $0.407$ & -- & $-0.156$ & $-0.053$ & --\tabularnewline
Diquark~\cite{Ke:2012wa} & $0.466$ & $0.736$ & -- & $-0.130$ & $-0.090$ & --\tabularnewline
\hline 
$\Xi_{cc}\to\Lambda_{c}$ & $f_{1}(0)$ & $f_{2}(0)$ & $f_{3}(0)$ & $g_{1}(0)$ & $g_{2}(0)$ & $g_{3}(0)$\tabularnewline
\hline 
This work & $0.517\pm0.071$ & $-0.036\pm0.007$ & -- & $0.155\pm0.019$ & $-0.072\pm0.012$ & --\tabularnewline
Diquark~\cite{Wang:2017mqp} & $0.790$ & $-0.008$ & -- & $0.224$ & $-0.050$ & --\tabularnewline
QCDSR~\cite{Shi:2019hbf}$^{*}$ & $0.63\pm0.20$ & $-0.05\pm0.02$ & $-0.81\pm0.26$ & $0.24\pm0.08$ & $-0.11\pm0.03$ & $-0.84\pm0.30$\tabularnewline
LCSR~\cite{Shi:2019fph} & $0.81\pm0.01$ & $0.32\pm0.01$ & $-0.90\pm0.07$ & $1.09\pm0.02$ & $-0.86\pm0.02$ & $0.76\pm0.01$\tabularnewline
NRQM~\cite{Perez-Marcial:1989sch} & $0.36$ & $0.14$ & $0.08$ & $0.20$ & $0.01$ & $-0.03$\tabularnewline
MBM~\cite{Perez-Marcial:1989sch} & $0.45$ & $0.01$ & $-0.28$ & $0.15$ & $0.01$ & $-0.70$\tabularnewline
\hline 
\end{tabular}
\end{table*}

\begin{table*}
\caption{Our decay widths (in units of $10^{-14}\ {\rm GeV}$) are compared
with other results in the literature. The asterisk on Ref.~\cite{Shi:2019hbf}
indicates that, in this literature, although we made a mistake in
the sign for the axial-vector form factors, the prediction for the
decay width is not affected. }
\label{Tab:comparision_decay_widths} %
\begin{tabular}{c|c|c|c|c|c}
\hline 
$\Lambda_{b}\to\Lambda_{c}e^{-}\bar{\nu}_{e}$ & Decay width & $\Sigma_{b}\to\Sigma_{c}e^{-}\bar{\nu}_{e}$ & Decay width & $\Xi_{cc}\to\Lambda_{c}e^{+}\nu_{e}$ & Decay width\tabularnewline
\hline 
This work & $2.54\pm0.33$ & This work & $0.870\pm0.061$ & This work & $0.755\pm0.219$\tabularnewline
Three-quark~\cite{Ke:2019smy} & $2.78$ & Three-quark~\cite{Ke:2019smy} & $1.03$ & Diquark~\cite{Wang:2017mqp} & $1.05$\tabularnewline
Diquark~\cite{Zhao:2018zcb} & $3.96$ & Diquark~\cite{Ke:2012wa} & $0.908$ & QCDSR~\cite{Shi:2019hbf}$^{*}$ & $0.76\pm0.37$\tabularnewline
Diquark ~\cite{Ke:2007tg} & $3.39$ &  &  & LCSR~\cite{Shi:2019fph} & $3.95\pm0.21$\tabularnewline
QCDSR~\cite{Zhao:2020mod} & $2.96\pm0.48$ &  &  &  & \tabularnewline
LQCD~\cite{Detmold:2015aaa} & $2.35\pm0.15$ &  &  &  & \tabularnewline
\hline 
\end{tabular}
\end{table*}

\section{Conclusions and discussions}

In this work, a three-quark picture is constructed using a bottom-up
approach for baryons in light-front quark model, where quark spinors
and Dirac matrices act as building blocks. The shape parameters, which
characterize the momentum distribution inside a baryon, are determined
with the help of the pole residue of the baryon. Some semileptonic decays are investigated under this three-quark picture. The relation between
the three-quark picture and the diquark picture is clarified. There
is still a small flaw worth pointing out, that is, when determining
the shape parameters, we demand that $\lambda_{1}\approx\lambda_{2}\approx\lambda_{{\rm QCDSR}}$,
some uncertainty can still be introduced. A better prescription we can
think of is to do global fitting. Given that our main goal in this
article is to develop a set of methods, such a more detailed consideration
is left for our future work. Here are some prospects. 
\begin{itemize}
\item At this point, we have constructed a relatively complete three-quark
picture for baryons, which can be applied
to study semileptonic, nonleptonic, strong,
and electromagnetic decay processes of heavy flavor baryons in the future.
\item When building the model, we found that Lorentz boost plays a
crucial role. As can be seen in Appendix~\ref{app:wave_functions}, in spin space, we first couple quark 3 and quark 2 to
form a ``diquark'', then couple this ``diquark'' to quark
1. Obviously, this bottom-up modeling approach can be generalized
to multiquark states. We may establish a unified theoretical framework
for describing multiquark states. 
\item As a by-product of model construction, we can easily obtain an
improved definition of baryon interpolating current. The
hadron interpolating currents are the starting point of Lattice
QCD and QCD sum rules, and therefore are of great importance.
These new interpolating currents can be applied to study more detailed problems,
such as the $\Xi_{Q}-\Xi_{Q}^{\prime}$ mixing~\cite{Sun:2023noo}. 
\end{itemize}

\section*{Acknowledgements}

The authors are grateful to Profs.~Hai-Yang Cheng, Chun-Khiang Chua, Run-Hui Li, Wei Wang and Drs.~Chia-Wei Liu, Xiao-Yu Sun, Zhi-Peng Xing, Jiabao Zhang for valuable discussions. This work is supported in part by National Natural Science Foundation of China under Grant No.~12065020. 
	
\appendix
	
\section{Spin wavefunctions}

\label{app:wave_functions}

In this Appendix, we will derive the spin wavefunctions of $\Lambda_{Q}$,
$\Sigma_{Q}$, and $\Sigma_{Q}^{*}$. Spinor and Dirac matrix conventions,
as well as some important equations, readers can refer to Ref.~\cite{Chua:2018lfa}.

In Eq.~(\ref{eq:baryon_state}),
$\Psi^{SS_{z}}$ is defined as
\begin{align}
 & \Psi^{SS_{z}}(\tilde{p}_{i},\lambda_{i})\nonumber \\
= & \sum_{s_{1}s_{2}s_{3}}\langle\lambda_{1}|{\cal R}_{M}^{\dagger}(\tilde{p}_{1},m_{1})|s_{1}\rangle\langle\lambda_{2}|{\cal R}_{M}^{\dagger}(\tilde{p}_{2},m_{2})|s_{2}\rangle\nonumber \\
 & \times\langle\lambda_{3}|{\cal R}_{M}^{\dagger}(\tilde{p}_{3},m_{3})|s_{3}\rangle\nonumber \\
 & \times\langle\frac{1}{2}\frac{1}{2};s_{3}s_{2}|S_{23}s_{23}\rangle\langle\frac{1}{2}S_{23};s_{1}s_{23}|SS_{z}\rangle\nonumber \\
 & \times\Phi(x_{i},k_{i\perp}),\label{eq:Psi_Melosh}
\end{align}
where $S_{23}$ is the total spin of the ``diquark'', and $S_{23}=0,1,1$
for $\Lambda_{Q}$, $\Sigma_{Q}$, $\Sigma_{Q}^{*}$, respectively.
An instant spinor $u_{D}(p,s)$ is Melosh transformed into a light-front
spinor $u(p,s)$ by
\begin{align}
 & \sum_{s_{i}}\langle\lambda_{i}|{\cal R}_{M}^{\dagger}(\tilde{p}_{i},m_{i})|s_{i}\rangle\bar{u}_{D}(p_{i},s_{i})\nonumber \\
= & \sum_{s_{i}}\bar{u}(p_{i},\lambda_{i})\frac{u_{D}(p_{i},s_{i})\bar{u}_{D}(p_{i},s_{i})}{2m_{i}}=\bar{u}(p_{i},\lambda_{i}).\label{eq:Melosh_spinor}
\end{align}
Once the Clebsch-Gordan (CG) coefficients in Eq.~(\ref{eq:Psi_Melosh})
are rewritten into the product of instant spinors and Dirac matrices, the instant spinors can then be transformed
into the light-front ones using Eq.~(\ref{eq:Melosh_spinor}). Therefore,
in the following, we will focus on rewriting the CG coefficients in
Eq.~(\ref{eq:Psi_Melosh}) into the product of instant spinors and Dirac matrices. Note
that the spinors appearing below are all instant spinors (we have
omitted their subscript $D$), except $u(\bar{P},S_{z})$ and $u_{\mu}(\bar{P},S_{z})$.
When one of these two spinors is involved, we always take the rest frame
of $\bar{P}$, where its instant form and light-front form coincide.

It is worth pointing out that the proof given here does not introduce any additional assumptions, for example,
it does not assume heavy quark symmetry,
nor does it depend on the coordinate system selection of LFQM.
In addition, one can clearly see that,
the Lorentz boost between the rest frame of ``diquark" and the rest frame of $\bar{P}$
plays a crucial role for the case involving an axial-vector ``diquark".
(Of course, for the case involving a scalar ``diquark", the Lorentz boost is trivial.)

\subsubsection{To derive the spin wavefunction of $\Lambda_{Q}$}

$\Lambda_{Q}$ has quark components $udQ$, in which $ud$ are considered
to form a $0^{+}$ diquark.
\begin{itemize}
\item Step 1, couple the spins of quark 3 and quark 2 to form a scalar ``diquark''
\begin{align}
I & \equiv\bar{u}(p_{3},s_{3})\frac{(\bar{\slashed P}+M_{0})}{2M_{0}}\gamma_{5}(-C)\bar{u}^{T}(p_{2},s_{2}),\nonumber \\
 & =\sqrt{(e_{2}+m_{2})(e_{3}+m_{3})}\chi_{s_{3}}^{\dagger}(i\sigma_{2})\chi_{s_{2}}^{*}\nonumber \\
 & =\sqrt{2(e_{2}+m_{2})(e_{3}+m_{3})}\langle\frac{1}{2}\frac{1}{2};s_{3}s_{2}|0s_{23}\rangle.
\end{align}
\item Step 2, calculate the trivial coupling 
\begin{align}
II & \equiv\bar{u}(p_{1},s_{1})u(\bar{P},S_{z})\nonumber \\
 & =\sqrt{2M_{0}(e_{1}+m_{1})}\chi_{s_{1}}^{\dagger}\chi_{S_{z}}\nonumber \\
 & =\sqrt{2M_{0}(e_{1}+m_{1})}\langle\frac{1}{2}0;s_{1}0|\frac{1}{2}S_{z}\rangle.
\end{align}
\end{itemize}
Therefore, for $\Lambda_{Q}$, the CG coefficients in Eq.~(\ref{eq:Psi_Melosh})
can be rewritten into
\begin{align}
 & \langle\frac{1}{2}\frac{1}{2};s_{3}s_{2}|0s_{23}\rangle\langle\frac{1}{2}0;s_{1}0|\frac{1}{2}S_{z}\rangle\nonumber \\
= & A_{0}\bar{u}(p_{3},s_{3})(\bar{\slashed P}+M_{0})(-\gamma_{5})C\bar{u}^{T}(p_{2},s_{2})\nonumber \\
 & \times\bar{u}(p_{1},s_{1})u(\bar{P},S_{z})
\end{align}
with
\begin{equation}
A_{0}=\frac{1}{4\sqrt{M_{0}^{3}(e_{1}+m_{1})(e_{2}+m_{2})(e_{3}+m_{3})}}.
\end{equation}

\subsubsection{To derive the spin wavefunction of $\Sigma_{Q}$}

$\Sigma_{Q}$ also has quark components $udQ$, in which $ud$ are
considered to form a $1^{+}$ diquark. Technically, it is much more
complicated to arrive at the spin wavefunction of $\Sigma_{Q}$. 
\begin{itemize}
\item Step 1, couple the spins of quark 3 and quark 2 in the rest frame
of $p_{23}$ (defined below) to form an axial-vector ``diquark''
\begin{align}
I^{\mu} & \equiv\bar{u}(p_{3},s_{3})\frac{(\bar{\slashed P}+M_{0})}{2M_{0}}\gamma_{\perp}^{\mu}(p_{23})\nonumber \\
 & \times(-C)\bar{u}^{T}(p_{2},s_{2})
\end{align}
with
\begin{align}
\gamma_{\perp}^{\mu}(p_{23}) & =\gamma_{\perp}^{\mu}(\bar{P})-\frac{M_{0}p_{23}^{\mu}+m_{23}\bar{P}^{\mu}}{M_{0}(e_{23}+m_{23})}\frac{\gamma_{\perp}(\bar{P})\cdot p_{23}}{m_{23}},\nonumber \\
p_{23} & =p_{2}+p_{3},\qquad m_{23}^{2}=p_{23}^{2},\nonumber \\
\gamma_{\perp}^{\mu}(\bar{P}) & =\gamma^{\mu}-\slashed vv^{\mu},\qquad v^{\mu}=\bar{P}^{\mu}/M_{0}.
\end{align}
$p_{23}^{\mu}$ is the four-momentum of the ``diquark'', and $e_{23}$
is its energy in the rest frame of $\bar{P}$ and $m_{23}$ is its
invariant mass.
$\gamma_{\perp}^{\mu}(p_{23})$ and $\gamma_{\perp}^{\mu}(\bar{P})$
are related by a Lorentz boost. It can be shown that
\begin{align}
I^{\mu} & =\sqrt{2(e_{2}+m_{2})(e_{3}+m_{3})}\langle\frac{1}{2}\frac{1}{2};s_{3}s_{2}|1s_{23}\rangle\nonumber \\
 & \times\epsilon^{*\mu}(p_{23},s_{23}).\label{eq:Imu}
\end{align}
Eq.~(\ref{eq:Imu}) can be verified by considering three specific cases:
$\mu=0$, $\mu=1,2$, and $\mu=3$. Take the $\mu=0$ case as an example:
\begin{align}
I^{0}= & \frac{1}{\sqrt{e_{3}+m_{3}}}\left(\begin{array}{cc}
(e_{3}+m_{3})\chi_{s_{3}}^{\dagger} & -\chi_{s_{3}}^{\dagger}\vec{\sigma}\cdot\vec{p}_{3}\end{array}\right)\nonumber \\
 & \left(\begin{array}{cc}
1 & 0\\
0 & 0
\end{array}\right)(-\frac{M_{0}e_{23}+m_{23}M_{0}}{M_{0}(e_{23}+m_{23})}\frac{-\gamma^{3}|\vec{p}_{23}|}{m_{23}})\nonumber \\
 & \left(\begin{array}{cc}
0 & i\sigma_{2}\\
i\sigma_{2} & 0
\end{array}\right)\nonumber \\
 & \frac{1}{\sqrt{e_{2}+m_{2}}}\left(\begin{array}{c}
(e_{2}+m_{2})\chi_{s_{2}}^{*}\\
-(\vec{\sigma}\cdot\vec{p}_{2})^{T}\chi_{s_{2}}^{*}
\end{array}\right)\nonumber \\
= & \sqrt{(e_{2}+m_{2})(e_{3}+m_{3})}\frac{|\vec{p}_{23}|}{m_{23}}\chi_{s_{3}}^{\dagger}(i\sigma_{3}\sigma_{2})\chi_{s_{2}}^{*}\nonumber \\
= & \sqrt{2(e_{2}+m_{2})(e_{3}+m_{3})}\langle\frac{1}{2}\frac{1}{2};s_{3}s_{2}|1s_{23}\rangle\nonumber \\
 & \times\epsilon^{*0}(p_{23},s_{23}).\label{eq:I0}
\end{align}
In Eq.~(\ref{eq:I0}), in the rest frame of $\bar{P}$, we choose
the ``diquark'' to move along the $z$-axis, then its four-momentum
$p_{23}^{\mu}=(e_{23},0,0,|\vec{p}_{23}|)$, polarization vector $\epsilon^{\mu}(0)=(|\vec{p}_{23}|,0,0,e_{23})/m_{23}$. 
\item Step 2, couple the ``diquark'' to quark 1
\begin{equation}
T\equiv I^{\mu}\times\bar{u}(p_{1},s_{1})\Gamma_{1,23\mu}u(\bar{P},S_{z})
\end{equation}
with
\begin{equation}
\Gamma_{1,23\mu}=\frac{\gamma_{5}}{\sqrt{3}}\left(\gamma_{\mu}-\frac{M_{0}+m_{1}+m_{23}}{M_{0}(e_{23}+m_{23})}\bar{P}_{\mu}\right).
\end{equation}
It can be shown that
\begin{align}
T & =2\sqrt{M_{0}(e_{1}+m_{1})(e_{2}+m_{2})(e_{3}+m_{3})}\nonumber \\
 & \times\langle\frac{1}{2}\frac{1}{2};s_{3}s_{2}|1s_{23}\rangle\langle\frac{1}{2}1;s_{1}s_{23}|\frac{1}{2}S_{z}\rangle.\label{eq:T}
\end{align}
Eq.~(\ref{eq:T}) can be verified by considering two specific cases:
$s_{23}=0$, and $s_{23}=\pm$. Take $s_{23}=0$ as an example:
\begin{align}
T= & \sqrt{2(e_{2}+m_{2})(e_{3}+m_{3})}\langle\frac{1}{2}\frac{1}{2};s_{3}s_{2}|1s_{23}\rangle\nonumber \\
 & \frac{1}{\sqrt{e_{1}+m_{1}}}\left(\begin{array}{cc}
(e_{1}+m_{1})\chi_{s_{1}}^{\dagger} & -\chi_{s_{1}}^{\dagger}\vec{\sigma}\cdot\vec{p}_{1}\end{array}\right)\nonumber \\
 & \frac{1}{\sqrt{3}}\left(\begin{array}{cc}
0 & 1\\
1 & 0
\end{array}\right)\Big[\frac{1}{m_{23}}(|\vec{p}_{23}|\gamma^{0}-e_{23}\gamma^{3})\nonumber \\
 & \quad-\frac{M_{0}+m_{1}+m_{23}}{M_{0}(e_{23}+m_{23})}\frac{|\vec{p}_{23}|}{m_{23}}M_{0}\Big]\nonumber \\
 & \sqrt{2M_{0}}\left(\begin{array}{c}
\chi_{S_{z}}\\
0
\end{array}\right)\nonumber \\
= & \sqrt{2(e_{2}+m_{2})(e_{3}+m_{3})}\langle\frac{1}{2}\frac{1}{2};s_{3}s_{2}|1s_{23}\rangle\nonumber \\
 & \times\frac{1}{\sqrt{3}}\sqrt{e_{1}+m_{1}}\sqrt{2M_{0}}\chi_{s_{1}}^{\dagger}\sigma_{3}\chi_{S_{z}}\nonumber \\
= & 2\sqrt{M_{0}(e_{1}+m_{1})(e_{2}+m_{2})(e_{3}+m_{3})}\nonumber \\
 & \times\langle\frac{1}{2}\frac{1}{2};s_{3}s_{2}|1s_{23}\rangle\langle\frac{1}{2}1;s_{1}s_{23}|\frac{1}{2}S_{z}\rangle.
\end{align}
\item Step 3, tensor simplify $T$
\begin{align}
 & \bar{u}(p_{3},s_{3})(\bar{\slashed P}+M_{0})\gamma_{\perp}^{\mu}(p_{23})(-C)\bar{u}^{T}(p_{2},s_{2})\nonumber \\
 & \times\bar{u}(p_{1},s_{1})\Gamma_{1,23\mu}u(\bar{P},S_{z})\nonumber \\
= & \bar{u}(p_{3},s_{3})(\bar{\slashed P}+M_{0})\gamma_{\perp}^{\mu}(\bar{P})(-C)\bar{u}^{T}(p_{2},s_{2})\nonumber \\
 & \times\bar{u}(p_{1},s_{1})\Gamma_{1,23\mu}u(\bar{P},S_{z})\nonumber \\
= & \bar{u}(p_{3},s_{3})(\bar{\slashed P}+M_{0})\gamma_{\perp}^{\mu}(\bar{P})(-C)\bar{u}^{T}(p_{2},s_{2})\nonumber \\
 & \times\bar{u}(p_{1},s_{1})\frac{\gamma_{5}}{\sqrt{3}}\gamma_{\mu}u(\bar{P},S_{z})\nonumber \\
= & \bar{u}(p_{3},s_{3})(\bar{\slashed P}+M_{0})(\gamma^{\mu}-v^{\mu})(-C)\bar{u}^{T}(p_{2},s_{2})\nonumber \\
 & \times\bar{u}(p_{1},s_{1})\frac{\gamma_{5}}{\sqrt{3}}\gamma_{\mu}u(\bar{P},S_{z})\nonumber \\
= & \bar{u}(p_{3},s_{3})(\bar{\slashed P}+M_{0})(\gamma^{\mu}-v^{\mu})C\bar{u}^{T}(p_{2},s_{2})\nonumber \\
 & \times\bar{u}(p_{1},s_{1})(\frac{1}{\sqrt{3}}\gamma_{\mu}\gamma_{5})u(\bar{P},S_{z}),
\end{align}
where, in the second step, we have used $\gamma_{\perp}^{\mu}(\bar{P})\cdot\bar{P}=0$. 
\end{itemize}
Therefore, for $\Sigma_{Q}$, the CG coefficients in Eq.~(\ref{eq:Psi_Melosh})
can be rewritten into
\begin{align}
 & \langle\frac{1}{2}\frac{1}{2};s_{3}s_{2}|1s_{23}\rangle\langle\frac{1}{2}1;s_{1}s_{23}|\frac{1}{2}S_{z}\rangle\nonumber \\
= & A_{1}\bar{u}(p_{3},s_{3})(\bar{\slashed P}+M_{0})(\gamma^{\mu}-v^{\mu})C\bar{u}^{T}(p_{2},s_{2})\nonumber \\
 & \times\bar{u}(p_{1},s_{1})(\frac{1}{\sqrt{3}}\gamma_{\mu}\gamma_{5})u(\bar{P},S_{z})
\end{align}
with
\begin{equation}
A_{1}=\frac{1}{4\sqrt{M_{0}^{3}(e_{1}+m_{1})(e_{2}+m_{2})(e_{3}+m_{3})}}.
\end{equation}

\subsubsection{To derive the spin wavefunction of $\Sigma_{Q}^{*}$}

$\Sigma_{Q}^{*}$ also has quark components $udQ$, in which $ud$
are considered to form a $1^{+}$ diquark.
\begin{itemize}
\item Step 1, same as that of $\Sigma_{Q}$.
\item Step 2, couple the ``diquark'' to quark 1
\begin{equation}
T^{\prime}\equiv I^{\mu}\times\bar{u}(p_{1},s_{1})\Gamma_{1,23\mu\nu}u^{\nu}(\bar{P},S_{z})
\end{equation}
with
\begin{equation}
\Gamma_{1,23\mu\nu}=-g_{\mu\nu}+\frac{\bar{P}_{\mu}(M_{0}p_{23\nu}+m_{23}\bar{P}_{\nu})}{M_{0}^{2}(e_{23}+m_{23})},
\end{equation}
and the vectorial spinor $u_{\mu}(\bar{P},S_{z})$ can be expressed
by~\cite{Auvil:1966eao,Zhao:2018mrg}
\begin{equation}
u_{\mu}(\bar{P},S_{z})=\sum_{s_{1},s_{23}}\langle\frac{1}{2}1;s_{1}s_{23}|\frac{3}{2}S_{z}\rangle u(\bar{P},s_{1})\epsilon_{\mu}(\bar{P},s_{23}).
\end{equation}
It can be shown that
\begin{align}
T^{\prime} & =2\sqrt{M_{0}(e_{1}+m_{1})(e_{2}+m_{2})(e_{3}+m_{3})}\nonumber \\
 & \times\langle\frac{1}{2}\frac{1}{2};s_{3}s_{2}|1s_{23}\rangle\langle\frac{1}{2}1;s_{1}s_{23}|\frac{3}{2}S_{z}\rangle.\label{eq:Tprime}
\end{align}
Eq.~(\ref{eq:Tprime}) can be proved as follows:
\begin{align}
 & \frac{T^{\prime}}{\sqrt{2(e_{2}+m_{2})(e_{3}+m_{3})}\langle\frac{1}{2}\frac{1}{2};s_{3}s_{2}|1s_{23}\rangle}\nonumber \\
= & \bar{u}(p_{1},s_{1})\Big[-\epsilon^{*\mu}(p_{23},s_{23})\nonumber \\
 & \quad+\frac{M_{0}p_{23}^{\mu}+m_{23}\bar{P}^{\mu}}{M_{0}^{2}(e_{23}+m_{23})}\epsilon^{*}(p_{23},s_{23})\cdot\bar{P}\Big]u_{\mu}(\bar{P},S_{z})\nonumber \\
= & \frac{1}{\sqrt{e_{1}+m_{1}}}\left(\begin{array}{cc}
(e_{1}+m_{1})\chi_{s_{1}}^{\dagger} & -\chi_{s_{1}}^{\dagger}\vec{\sigma}\cdot\vec{p}_{1}\end{array}\right)\nonumber \\
 & \times(-\epsilon^{*\mu}(\bar{P},s_{23}))\nonumber \\
 & \times\sum_{\lambda_{1}\lambda_{2}}\langle\frac{1}{2}1;\lambda_{1}\lambda_{2}|\frac{3}{2}S_{z}\rangle u(\bar{P},\lambda_{1})\epsilon_{\mu}(\bar{P},\lambda_{2})\nonumber \\
= & \frac{1}{\sqrt{e_{1}+m_{1}}}\left(\begin{array}{cc}
(e_{1}+m_{1})\chi_{s_{1}}^{\dagger} & -\chi_{s_{1}}^{\dagger}\vec{\sigma}\cdot\vec{p}_{1}\end{array}\right)\nonumber \\
 & \times\sum_{\lambda_{1}\lambda_{2}}\langle\frac{1}{2}1;\lambda_{1}\lambda_{2}|\frac{3}{2}S_{z}\rangle\sqrt{2M_{0}}\left(\begin{array}{c}
\chi_{\lambda_{1}}\\
0
\end{array}\right)\delta_{s_{23},\lambda_{2}}\nonumber \\
= & \sqrt{e_{1}+m_{1}}\sqrt{2M_{0}}\sum_{\lambda_{1}\lambda_{2}}\langle\frac{1}{2}1;\lambda_{1}\lambda_{2}|\frac{3}{2}S_{z}\rangle\delta_{s_{1},\lambda_{1}}\delta_{s_{23},\lambda_{2}}\nonumber \\
= & \sqrt{2M_{0}(e_{1}+m_{1})}\langle\frac{1}{2}1;s_{1}s_{23}|\frac{3}{2}S_{z}\rangle.
\end{align}
Here, we have used 
\begin{align}
\epsilon^{\mu}(p_{23},s_{23})-\frac{M_{0}p_{23}^{\mu}+m_{23}\bar{P}^{\mu}}{M_{0}^{2}(e_{23}+m_{23})}\epsilon(p_{23},s_{23})\cdot\bar{P}\nonumber \\
=\epsilon^{\mu}(\bar{P},s_{23})
\end{align}
 and 
\begin{equation}
\epsilon^{*\mu}(\bar{P},s_{23})\epsilon_{\mu}(\bar{P},\lambda_{2})=-\delta_{s_{23},\lambda_{2}}.
\end{equation}
\item Step 3, tensor simplify $T^{\prime}$
\begin{align}
 & \bar{u}(p_{3},s_{3})(\bar{\slashed P}+M_{0})\gamma_{\perp}^{\mu}(p_{23})(-C)\bar{u}^{T}(p_{2},s_{2})\nonumber \\
 & \times\bar{u}(p_{1},s_{1})\Gamma_{1,23\mu\nu}u^{\nu}(\bar{P},S_{z})\nonumber \\
= & \bar{u}(p_{3},s_{3})(\bar{\slashed P}+M_{0})\gamma_{\perp}^{\mu}(\bar{P})(-C)\bar{u}^{T}(p_{2},s_{2})\nonumber \\
 & \times\bar{u}(p_{1},s_{1})\Gamma_{1,23\mu\nu}u^{\nu}(\bar{P},S_{z})\nonumber \\
= & \bar{u}(p_{3},s_{3})(\bar{\slashed P}+M_{0})\gamma_{\perp}^{\mu}(\bar{P})(-C)\bar{u}^{T}(p_{2},s_{2})\nonumber \\
 & \times\bar{u}(p_{1},s_{1})(-g_{\mu\nu})u^{\nu}(\bar{P},S_{z})\nonumber \\
= & \bar{u}(p_{3},s_{3})(\bar{\slashed P}+M_{0})(\gamma^{\mu}-v^{\mu})C\bar{u}^{T}(p_{2},s_{2})\nonumber \\
 & \times\bar{u}(p_{1},s_{1})u_{\mu}(\bar{P},S_{z}).
\end{align}
\end{itemize}
Therefore, for $\Sigma_{Q}^{*}$, the CG coefficients in Eq.~(\ref{eq:Psi_Melosh})
can be rewritten into
\begin{align}
 & \langle\frac{1}{2}\frac{1}{2};s_{3}s_{2}|1s_{23}\rangle\langle\frac{1}{2}1;s_{1}s_{23}|\frac{3}{2}S_{z}\rangle\nonumber \\
= & A_{1}^{\prime}\bar{u}(p_{3},s_{3})(\bar{\slashed P}+M_{0})(\gamma^{\mu}-v^{\mu})C\bar{u}^{T}(p_{2},s_{2})\nonumber \\
 & \times\bar{u}(p_{1},s_{1})u_{\mu}(\bar{P},S_{z})
\end{align}
with
\begin{equation}
A_{1}^{\prime}=\frac{1}{4\sqrt{M_{0}^{3}(e_{1}+m_{1})(e_{2}+m_{2})(e_{3}+m_{3})}}.
\end{equation}
One can see that $A_{0}=A_{1}=A_{1}^{\prime}$.

\end{document}